\documentclass[sigconf,10pt]{acmart}

\usepackage{fancyhdr}
 \usepackage{amsmath}
\usepackage{cases}
\usepackage{stfloats}
\usepackage{enumerate}
\usepackage{url}
\usepackage{graphicx}
\usepackage{wrapfig}
\usepackage{picinpar}
\usepackage{cutwin}
\usepackage{picins}
\usepackage{balance}
\usepackage{algorithm}
\usepackage{algpseudocode}

 \usepackage{etoolbox}

\newtoggle{ACM}
\toggletrue{ACM}

\newtoggle{shortV}
\newtoggle{longV}
\toggletrue{shortV}
\toggletrue{longV}
 \iftoggle{shortV}{
 \togglefalse{longV}
 }

\newtoggle{IEEEcls}
\toggletrue{IEEEcls}
\togglefalse{IEEEcls}

\newtoggle{ElsJ}
\toggletrue{ElsJ}
\togglefalse{ElsJ}

\floatname{algorithm}{Algorithm}

\iftoggle{IEEEcls}{
\usepackage{amsthm}
\theoremstyle{plain}
\newtheorem{theorem}{Theorem}
\newtheorem{corollary}{Corollary}
\theoremstyle{definition}
\newtheorem{definition}{Definition}
\newtheorem{property}{Proposition} 
\newtheorem{lemma}{Lemma}

}

\iftoggle{ElsJ}{
\usepackage{amsthm}
\theoremstyle{plain}

\theoremstyle{definition}
\newtheorem{definition}{Definition}

}

\newif\ifNotUse  
 \NotUsetrue
 
 \newif\ifNotUse  
 \NotUsetrue

 \iftoggle{ACM}{
\setcopyright{none}  
\acmDOI{}
\acmPrice{}
\acmISBN{}
}
 
\begin{document}

\title{Distributed super point cardinality estimation under sliding time window for high speed network}

 \iftoggle{ElsJ}{
\author[seu_cs]{Jie Xu\corref{cor1}}
\ead{xujieip@163.com}

\cortext[cor1]{Corresponding author}
\address[seu_cs]{School of Computer Science and Engineering, South East University, Nanjing, China}
\address[seu_math]{School of Mathematics, South East University, Nanjing, China}
}

\author{Jie Xu}
\affiliation{%
  \institution{School of computer science and engineer, Southeast university}
  \city{Nanjing}
  \state{China}
 }
\email{xujieip@163.com}

\begin{abstract}
Super point is a special kind of host whose cardinality, the number of contacting hosts in a certain period, is bigger than a threshold. Super point cardinality estimation plays important roles in network field. This paper proposes a super point cardinality estimation algorithm under sliding time window. To maintain the state of previous hosts with few updating operations, a novel counter, asynchronous time stamp (AT), is proposed. For a sliding time window containing k time slices, AT only needs to be updated every k time slices at the cost of 1 more bit than a previous state-of-art counter which requires $log_2(k+1)$ bits but updates every time slice. Fewer updating operations mean that more AT could be contained to acquire higher accuracy in real-time. This paper also devises a novel reversible hash function scheme to restore super point from a pool of AT. Experiments on several real-world network traffic illustrate that the algorithm proposed in this paper could detect super points and estimate their cardinalities under sliding time window in real time.
\end{abstract} 

 \iftoggle{ElsJ}{
\begin{keyword}
super points detection \sep distributed computing \sep GPU computing \sep network measurement
\end{keyword}
}
  \iftoggle{ACM}{
\keywords{super point detection, cardinality estimation, sliding time window, GPU computing} 
 }
 
\maketitle

\section{Introduction}
 Suppose there are two networks ANet and BNet. These two networks are contacting with each other through an edge router ER. ANet might be a city-wide network or even a country-wide network. And BNet might be another citywide network or the Internet. All traffic between ANet and BNet could be observed from ER. For a host "aip" in ANet, the number of hosts in BNet which sending packets to or receiving packets from it in a certain period is called aip's cardinality. When aip's cardinality is more than a threshold $\theta$, aip is called a super point\cite{hsd:detectionSuperSourcesInHighSpeed}. 

Super point is a special kind of hosts which relates to many network events\cite{HSD:CardinalityChangebasedEarlyDetectionLargescaleCyberattacks}, such as DDoS\cite{HSD:SuperpointbasedDetectionAgainstDistributedDenialServiceDDoSFloodingAttacks}, network scanning\cite{scan:surveyPortScansAndDetection}  and so on. Host's cardinality is also an important character in network management and measurement. Calculating and monitoring super point's cardinality is an efficient way for high-speed network real-time management because the super point takes up only a fraction of the total hosts.  This topic has been received great attention for a long time, and many excellent algorithms have been proposed in recent years. 

The "period" in the super point definition could be a discrete time window or a sliding time window\cite{SDC:IMC2003:IdentifyingFrequentItemsSlidingWindowsOnlinePacketStreams}. Most existing algorithms only work under discrete time window, in which there is no duplicating period between two adjacent windows.

Super point's cardinality estimation under discrete time window is simple because it doesn't need to maintain hosts' state in the previous time slices. But the estimating result has the following two problems:
\begin{enumerate}
\item The result is affected by the starting of a discrete time window, and it fails to detect and estimate the super point which spans the boundary of two adjacent windows.
\item The result is reported with high latency. 
\end{enumerate}
This two weakness of discrete time window comes from its moving step. The moving step of discrete time window equals its size. The bigger monitor period, the higher latency and more errors. Sliding time window solves these two problems together because its moving step has no relation to its window size. But super point cardinality estimation under sliding time window is more complex than that under discrete time window because it maintains hosts state of previous time and estimates super point's cardinality more frequently. 

This paper proposes a super point cardinality estimation algorithm under sliding time window. To maintain the state of previous hosts with few updating operations, a novel counter, asynchronous time stamp(AT), is proposed. For a sliding time window containing k time slices, AT only need to be updated every k time slices at the cost of 1 more bit than a previous state-of-art counter\cite{Iwqos2017:ACE:PerflowCountingForBigNetworkDataStreamOverSlidingWindows} which requires $log_2(k+1)$ bits but updates every time slice. Fewer updating operations mean that more counters could be contained to acquire higher accuracy in real-time. This paper also devises a novel reversible hash function scheme which is the key to restore super points. Based on asynchronous timestamp and this reversible hash function scheme, a novel sliding super point cardinality estimation algorithm, ASSE, is proposed.
ASSE is also an available parallel algorithm which could be deployed on GPU for nowadays high bandwidth network. The main contribution of this paper is listed below.
\begin{enumerate}
\item Devise a novel counter to record host state under sliding time window. This counter only needs to be updated every k time slices at the cost of one more bit.
\item Design a high random reversible hash function scheme. It maps an IP to several random values. And this IP could also be restored from these hashed values. It plays an important role in super point detection.
\item Propose a new super point cardinality estimating algorithm under sliding time window which uses fix number of the novel counters.
\item Deploy the sliding super point cardinality estimation algorithm on a common GPU to deal with a core network in real time.
\end{enumerate}

In the next section, we will introduce related works. In section 3, a novel cardinality estimating algorithm which works under sliding time window is proposed. Section 4 introduces the novel reversible hash function scheme and our super point cardinality estimating algorithm under sliding time window. In this section, we also introduced how to deploy our algorithm on GPU. Section 5 shows experiments of real-world core network traffics. And we conclude in the last section.

\section{Related work}
\subsection{Problem definition} 
Measuring core network's properties, such as traffic size, packets number, host cardinality and so on, is the foundation of network management. There are huge hosts in a core network. But only a small fraction of them have great influence on the network performance. This paper focuses on how to detect a kind of special hosts from the perspective of cardinality over sliding time window. Suppose there is a core network, $ANet$, which is under the management of some organizations, institutes or ISP(internet service provider). $ANet$ communicates with other networks, denoted as $BNet$, through a set of edge routers ER. For a host $aip \in ANet$, its cardinality is the number of hosts in $BNet$ which communicate with $aip$ through $ER$ in a time window. When the cardinality of $aip$ is greater than or equal to a threshold $\theta$, $aip$ is a super point. The managers of $ANet$ have the authority to inspect every packet between $ANet$ and $BNet$ through ER.  So the task of super point cardinality estimation is to detect super points and estimate their cardinalities by scanning all packets passing through $ER$. Cardinality estimation and super point detection are hot topics in network research.
\subsection{Cardinality estimation}
For a host $aip \in ANet$, let $Pkt(aip, t_0, t_1)$ represent the stream of packets passing through $ER$ from time point $t_0$ to $t_1$ whose source or destination IP address is $aip$. The period from time point $t_0$ to $t_1$ is the time window, written as $TW(t_0, t_1)$. An IP pair which is similar to $<aip,bip>$ could be extracted from each packet in $Pkt(aip, t_0, t_1)$ where $bip$ is the other host in the packet. We also call $bip$ the opposite host of $aip$. Let $IPair(aip, t_0, t_1)$ represent to the stream of IP pairs corresponding to $Pkt(aip, t_0, t_1)$. Because a host $bip \in BNet$ could send several packets to or receive several packets from $aip$ in a time window, IP pair $<aip, bip>$ can appear many times in $IPair(aip, t_0, t_1)$. The number of distinct IP pairs in $IPair(aip, t_0, t_1)$ is the cardinality of $aip$. Let $OP(aip, t_0, t_1)$ represent the set of hosts in $BNet$ that communicate with $aip$ from time point $t_0$ to $t_1$ and $|OP(aip, t_0, t_1)|$ represent the number of hosts in $OP(aip, t_0, t_1)$. Estimating the cardinality of $aip$ in $W(t_0, t_1)$ is to calculate $|OP(aip, t_0, t_1)|$ by scanning $IPair(aip, t_0, t_1)$. 

Many cardinality estimation algorithms have been proposed. Cardinality estimation algorithms use fix number of the counter to record and calculate the cardinality of a host. All these algorithms use a counter vector containing $g$ counters. What is preserved in a counter, how to update counters and how to estimate the cardinality from the counter vector are special in different algorithms.

Flajolet et al. \cite{PCSA:ProbabilisticCountingAlgorithmsForDataBaseApplications}firstly proposed such an algorithm which is called Probabilistic Counting with Stochastic Averaging, PCSA. Each counter in PCSA is a bitmap containing 32 bits. For every opposite host of $aip$, a random selecting counter is used to record the least significant bit of this element. Least significant bit, LSB, is the first `1' bit starting from the right. After scanning all elements in the stream, the value of each counter is its least zero position starting from the right. Cardinality could be acquired according to the sum of every counter. Scheuermann et al. \cite{PCSA_correct:NearOptimalCompressionProbabilisticCountingSketchesNetworkingApplications} proposed a more accuracy estimating equation when the load factor is smaller than 20. Load factor is the ratio of cardinality to s. 

The task of every counter in PCSA is to record the lowest zero position of every element. For an IPv4 address, the biggest value of least zero position is 32. But PCSA uses 32 bits to record the least zero position which leaves great improvement space. Because the biggest value of each counter is 32, 5 bits are big enough to represent it. Motivated by this idea, Philippe et al. proposed the LogLog counting algorithm\cite{DC:LoglogCountingOfLargeCardinalitiesDurand2003}. Unlike PCSA, each counter of LogLog records the leftmost `1' bit position of every element in the stream. Loglog estimates the cardinality according to the geometric mean value of all counters. 
Many algorithms are derived from LogLog. Flajolet et al.\cite{DC:HyperLogLogTheAnalysisOfANearoptimalCardinalityEstimationAlgorithm} found that when using the harmonic mean value of all the counters, the accuracy will be improved. And their proposed HyperLogLog algorithm based on this idea. MinCount\cite{DC2009:OrderStatisticsEstimatingCardinalitiesMassiveDataSets} is another algorithm similarly to LogLog. But it hashes every opposite host to a real value between [0,1] uniformly, and every counter stores the minimum of hashed value it has ever seen. The size of every counter could be adjusted for different precision. 

\begin{figure}[!ht]
\centering
\includegraphics[width=0.47\textwidth]{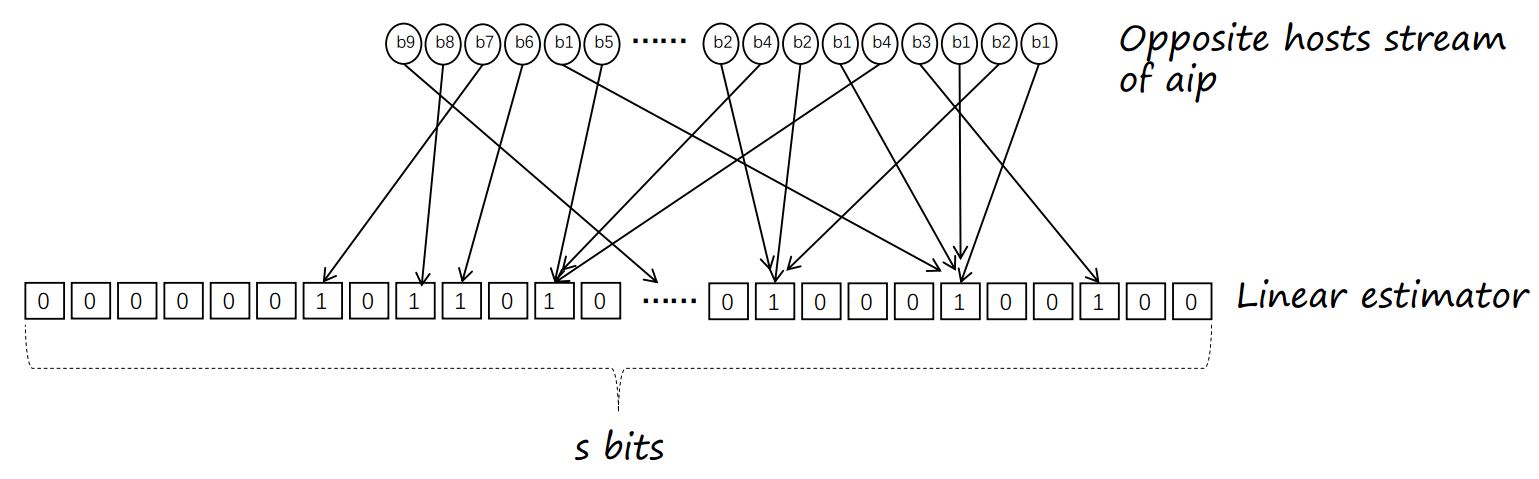}
\caption{Linear estimator}
\label{Linear_estimator_illustrate}
\end{figure}

Although these algorithms are memory efficient for big cardinality estimation, their accuracy is limited. Whang et al. \cite{DC:aLinearTimeProbabilisticCountingDatabaseApp} proposed a high accuracy cardinality estimation algorithm, Linear Estimator LE, based on maximum likelihood estimation. A counter in LE is a bit. LE uses a bit to record the appearance of opposite hosts. All of these bits are initialized to zero at the beginning. For every element in the stream, a randomly selected bit will be set to 1 as shown in figure \ref{Linear_estimator_illustrate}. Opposite hosts stream of $aip$ is the stream of hosts in $BNet$ extracted from $IPair(aip, t_0, t_1)$ by removing $aip$ of each IP pair. Every opposite host will be mapped to a bit by a random hash function\cite{hash_UniversalClassesOfHashFunctions}. After scanning all elements in a time window, LE estimates the cardinality based on the zero number in the bit vector. Suppose $g_0$ is the number of `0' bits. The cardinality of $aip$ will be estimated by equation \ref{eq_linearEstimator}.
\begin{equation}
\label{eq_linearEstimator}
|OP(aip,t_0,t_1)|=-g*\frac{g_0}{g}
\end{equation}

The accuracy of a cardinality estimating algorithm is evaluated by its standard error\cite{TON2017_CardinalityEstimationElephantFlowsACompactSolutionBasedVirtualRegisterSharing}. Let $n$ represent the cardinality of $aip$ and $\hat{n}$ is the estimating value acquired by an algorithm. The standard error of an algorithm is the standard error of $\frac{\hat{n}}{n}$, written as $\sigma$. Table \ref{tbl_cardinalityEstimating_compare} shows the accuracy and memory consumption of different algorithms when $\sigma=1\%$ and $n=5000$. 
\begin{table}
\centering
\caption{Different cardinality estimator compare}
\label{tbl_cardinalityEstimating_compare}
\begin{tabular}{c}                                                                                                                                                                                                                           
\centering
\includegraphics[width=0.45\textwidth]{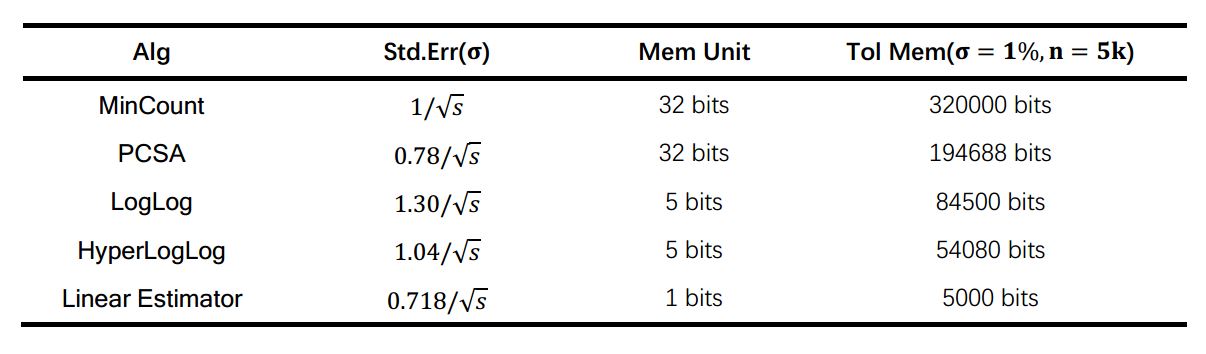}
\end{tabular}
\end{table}
From table \ref{tbl_cardinalityEstimating_compare}, we can see that LE uses the smallest memory to acquire the same accuracy as other algorithms. In this paper, we estimate the cardinality of super point under sliding time window by a novel estimator derived from LE. So our algorithm has the same accuracy as high as LE. 

\subsection{Sliding time window vs. discrete time window}
\begin{figure}[!ht]
\centering
\includegraphics[width=0.47\textwidth]{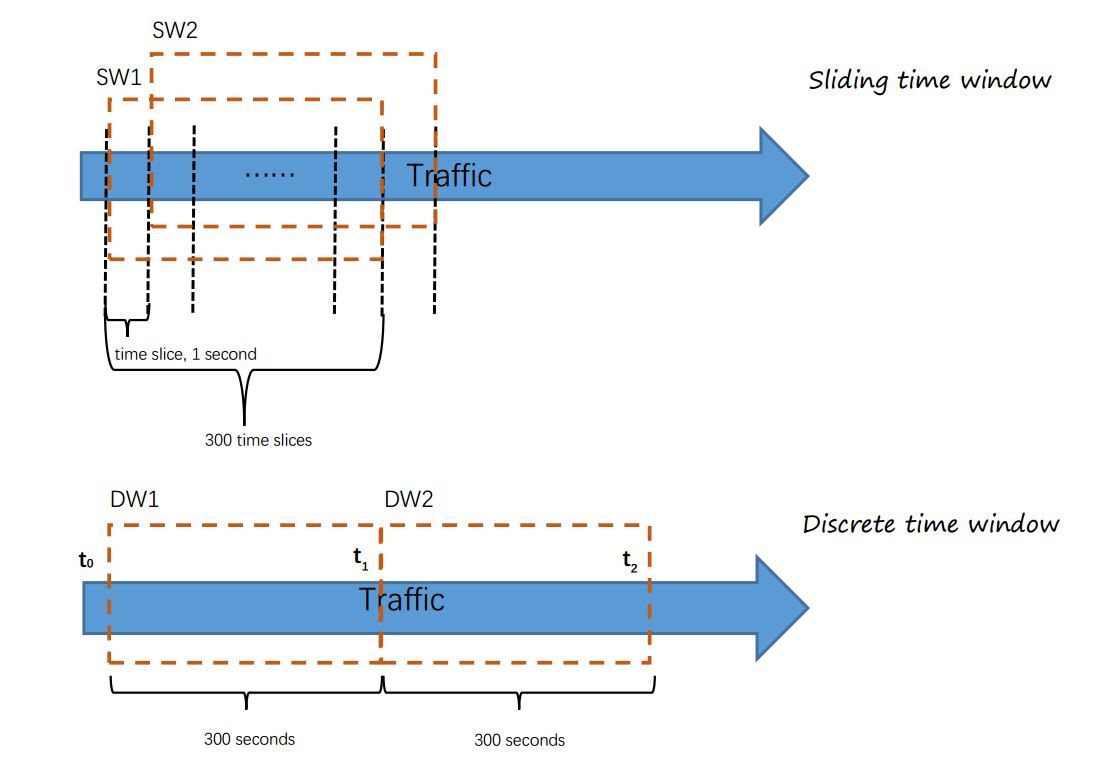}
\caption{Sliding time window and discrete time window}
\label{SlidingDiscreting_time_window}
\end{figure}

 Discrete time window and sliding time window are two kinds of the period for cardinality estimating as shown in figure \ref{SlidingDiscreting_time_window}. 

Traffic between network ANet and BNet could be divided into successive time slices which have the same duration. The length of a time slice could be 1 second, 1 minute or any period in different situations. A sliding time window, denoted as $W(t, k)$, contains k successive time slices starting from time point t as shown in the top part of figure \ref{SlidingDiscreting_time_window}. Sliding time window will move forward one time slice a time. So two adjacent sliding time windows contain k-1 same slices. When k is set to 1, there is no duplicate period between two adjacent windows, which is the case of the discrete time window in the bottom part of figure \ref{SlidingDiscreting_time_window}. In figure \ref{SlidingDiscreting_time_window}, the size of the time slice is set to 1 second for sliding time window and 300 seconds for the discrete time window. A sliding window in figure \ref{SlidingDiscreting_time_window} contains 300 time slices. In figure \ref{SlidingDiscreting_time_window}, the size of a sliding time window is equal to that of a discrete time window. 

Cardinality estimation under discrete time window is easy because it doesn't need to maintain the appearance of opposite hosts in another time window. But the result is affected by the starting of the discrete time window. When a super point has different opposite hosts in two adjacent time windows, it may be neglected under discrete time window. 

\iftoggle{longV}{
For example, suppose that $DW1$ starts from time point $t_0$ to time point $t_2$ and $DW2$ starts from time point $t_2$ to time point $t_4$ in figure \ref{SlidingDiscreting_time_window}. Let $t_1$ and $t_3$ be two time points in $DW1$ and $DW2$ separately and $TW(t_1, t_3)=300$ seconds. If $|OP(aip, t_1, t_2)|=512$ and $|OP(aip, t_2, t_3)|=512$, $aip$ is a super point in $TW(t_1, t_3)$. But $aip$ will never be detected out in $DW1$ nor $DW2$. By surveying a real-world 40Gb/s network, we found that discrete time window will lose average 14 such super points. Cardinality estimation algorithms discussed in previous subsection all work under discrete time window.
}

Sliding time window has higher accuracy than discrete time window because it monitors traffic in a much more scalable way\cite{SDC2010SHLL:SlidingHyperLogLogEstimatingCardinalityDataStreamOverSlidingWindow}. 
\iftoggle{longV}{
There are two kinds of sliding window: sliding time window and sliding packet window\cite{SDC2010SHLL:SlidingHyperLogLogEstimatingCardinalityDataStreamOverSlidingWindow}. Sliding time window splits traffic into equal size time slices and slides forward a time slice once a time. Sliding packet window contains a constant number of packets and slides forward a packet once a time\cite{Iwqos2017:ACE:PerflowCountingForBigNetworkDataStreamOverSlidingWindows}. Sliding packet window may work well in a small network, but not in a core network. For a core network, such as 40Gb/s city-wide network, there are millions of packets passing through the edge router. It means that there is a packet passing through in hundreds of nanoseconds. Reporting super point's cardinality over such a short period is difficult and unnecessary. In this paper, we focus on cardinality estimation under sliding time window.
}
Being required to preserve the state of opposite hosts in previous time slices, cardinality estimation under sliding time window is more burdensome. But many works have been down trying to solve this problem. The main idea is to replace each counter used in discrete time window with a more powerful structure which can tell if itself is active in the current time window. For a counter, if it is updated in $W(t,k)$, it is called active in this time window.

Fusy et al. \cite{SDC2007:EstimatingNumberActiveFlowsDataStreamOverSlidingWindow} extended MinCount to sliding window by maintaining a list of hosts that may become a minimum in a future window. The new algorithm is called Sliding MinCount. The minimum host is the latest arrived hosts among the set of hosts whose hashed value realizes the minimum in a sliding time window. When the time window sliding, Sliding MinCount updates every list and removes inactive hosts from these hosts list. But Sliding MinCount requires much space to store the minimum value of different time slices. In the worst case, each counter will maintain k minimum values in a time sliding window with k time slices. When using 32 bits to represent a minimum value, each counter of Sliding MinCount requires 32*k bits.

Chabchoub et al. \cite{SDC2010SHLL:SlidingHyperLogLogEstimatingCardinalityDataStreamOverSlidingWindow} replaced each counter in HyperLogLog with a list of future possible maxima(LFPM). Each cell of LFPM uses 4 bytes to store timestamp and 1 byte to store the max leftmost 1-bit. In a time sliding window with k time slices, LFPM contains $ln(n/s)$ cells on average. So the size of a LFPM is $40*ln(n/s)$ bits.  

Considering the high accuracy of LE, many algorithms are devised based on it. Kim et al. \cite{GLOBECOM2003_CountingNetworkFlowsInRealTime} used a time stamp vector, TSV, to replace the bit vector in LE. Every time stamp contains 64 bits. TSV can give the cardinality at any time for any size of the time window. But in practice, we don't need to query host's cardinality in such a way. For a window with k time slices, the size of each counter could be as small as $log_2(k)$ bits. Y. Zhou et al. \cite{Iwqos2017:ACE:PerflowCountingForBigNetworkDataStreamOverSlidingWindows} used an aging counter estimator (ACE) to delete old inactive counters in an approximate fashion. Unlike TSV, every timeout counter in ACE only requires $log_2(k+1)$ bits. It could be seen as divided into two processes: the first updates the vector for each IP pair, while the second is in charge of decreasing the timeout counters at the end of every time slice. If a timeout counter is visited by some host in a time slice, it will be set to the max value c. At the end of a time slice, if a timeout counter is not zero, it will be decreased by 1. If a timeout counter's value is no less than c-k-1, this counter is active in the sliding time window. Although ACE uses smaller memory than TSV does, it needs to updates every counter at the end of every time slice. 

Shan et al. devices an LRU-Sketch by combining a bitmap sketch with the least-recently-used (LRU) replacement policy. Each bit is replaced with an LRU structure which contains a head pointer, a tail pointer, and a time difference counter. In another word, LRU-Sketch is a double direction list. In any time slice, only the left-most entry of LRU-Sketch may become inactive. So the eviction of inactive entry takes constant time (O(1)) at the end of each time slice. But to maintain the feature of LRU-Sketch, it has to shift a node in LRU-Sketch to the tail every time it processes a packet. Changing a node's position in double direction list is very expensive which at least needs to modify four nodes' pointers. And these pointers of LRU forbid LRU-Sketches to share nodes between different hosts like vHLL\cite{TON2017_CardinalityEstimationElephantFlowsACompactSolutionBasedVirtualRegisterSharing}. So LRU-Sketch could only estimate the cardinality of a single host. If we want to estimate all hosts' cardinalities in a core network, we have to assign a private LRU-Sketch for each host. It is inefficient for multi hosts' cardinalities estimation.

This paper propsed a memory efficient and few preserving requirement estimator, asynchronous timestam vector(ATV). Table \ref{tbl_slidingCE_compare} compare the difference of these state-of-the-art algorithms and ATV. All of these sliding cardinality estimators are based on some classic ones shown in "Basic Alg" column. Column "Opt" shows the operation complexity of different algorithms to preserve the state their structure.   

\begin{table}
\centering
\caption{Different sliding time cardinality estimators compare}
\label{tbl_slidingCE_compare}
\begin{tabular}{c}                                                                                                                                                                                                                           
\centering
\includegraphics[width=0.45\textwidth]{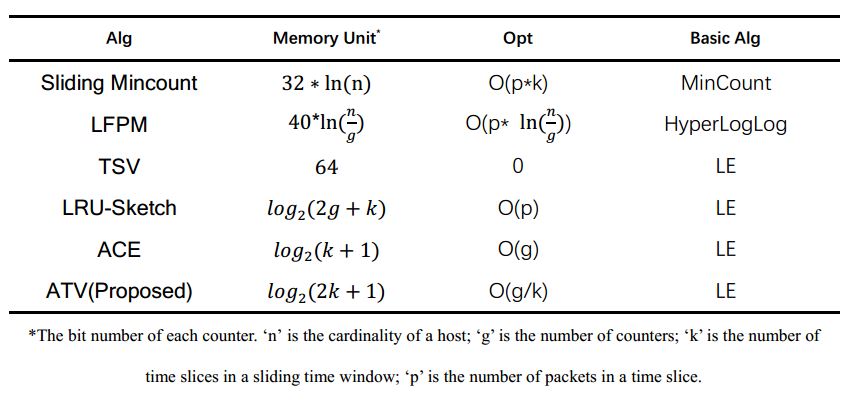}
\end{tabular}
\end{table}

ATV uses one more bit than ACE does. But the number of counters updated by ATV in a time slice is a factor of k smaller than that of ACE. With this merit, ATV can contain huge counters to acquire high accuracy.

\subsection{Multi hosts cardinality estimation}
In the core network, there are huge hosts. A precisely way to acquire all of these hosts' cardinalities is to allocate an estimator for each of them. But this way is memory wasting and slow. Recent algorithms use a fixed number of estimators to maintain and calculate all hosts' cardinalities. These algorithms could be classified into two branches: virtual estimator vector based and estimators array based.
Virtual counter based algorithms assign a logical counter vector for every host. Every host's logical counter vector shares counter with other's in a counter pool. The estimator of a virtual estimator could be LE \cite{HSD:SpreaderClassificationBasedOnOptimalDynamicBitSharing}\cite{HSD:GPU:2014:AGrandSpreadEstimatorUsingGPU}, HyperLogLog\cite{TON2017_CardinalityEstimationElephantFlowsACompactSolutionBasedVirtualRegisterSharing} and ACE \cite{Iwqos2017:ACE:PerflowCountingForBigNetworkDataStreamOverSlidingWindows}. But the result of the virtual estimator is affected by the number of hosts. When there are many hosts, counters in a virtual estimator will be over shared.

To reduce the affection of estimator sharing, algorithms based on estimator array which use $n$ estimators to estimate a host's cardinality at the same time are proposed. The main data structure is an array of estimators with n rows and m columns. For a host $aip$, each row randomly selects an estimator to record its cardinality. Linear estimator array is the most popular. Wang et al. \cite{HSD:ADataStreamingMethodMonitorHostConnectionDegreeHighSpeed} used a $3$ rows LE array (DCDS), and the number of columns of each row is a prime different from each other. Liu et al. \cite{HSD:DetectionSuperpointsVectorBloomFilter} used an $5*2^{12}$ LE array (VBF) to estimate hosts' cardinalities. DCDS and VBF map a host to each row's LE by Chinese remainder theorem(CRT) and sub bits of IP address separately. 

Because the high accuracy of estimator array, this paper designs an ATV array to estimate hosts cardinality under sliding time window.

\subsection{Super point detection}
There are huge hosts in a core network. 
\iftoggle{longV}{
For example, there are more than 1.2 million hosts appears on a 40 Gb/s network in 5 minutes (CERNET \cite{expdata:IPtraceCernetJS}, Oct 23, 2017). Estimating cardinalities of all these hosts is time wasting.
}
According to the research of Lan et al. \cite{NW2006：AMeasurementStudyOfCorrelationsOfInternetFlow}, only a small fraction of hosts is important for the network management. From the perspective of cardinality, we focus on super points whose cardinality is more than $\theta$. 
\iftoggle{longV}{
In the same traffic as the previous example, when $\theta=1024$, there are only 600 super points. If we allocate more monitoring resource to these crucial hosts, we will manage core network more efficiently.
}
But how to detect super points from huge hosts is a hard task. 

Virtual counter based algorithms can't restore super points because they didn't maintain the relationship between hosts and virtual counters. Unlike virtual counter based algorithms, estimator array based algorithms preserve the cardinality in $n$ estimators. In another word, a super point $sip$ corresponds to a tuple of estimator indexes, $r(sip)=\{<r_0(sip), r_1(sip),$ $\cdots, r_{n-1}(sip)>\}$. If we can restore $sip$ from $r(sip)$, we would detect super points from this estimator array. Such kind of hash functions is called reversible hash functions.

Schweller et al. \cite{RSL2012:TheLogLogCountingReversibleSketchADistributedArchitectureForDetectingAnomaliesBackboneNetworks} proposed a reversible hash function, Reversible sketches($RS$). $RS$ firstly encode $sip$ to a random value $hip$ by function $f(sip)=a*sip\ mod\ p$ where $p$ is a prime number bigger than $2^{32}$ and $a$ is random number smaller than $p$. Notice that $sip$ could be decoded from $hip$ by function $f^{-1}(hip)=a^{-1}*hip\ mod\ p$ where $a^{-1} * a\ =1\ mod\ p$. This process is called IP mangling. $RS$ hashes $hip$ to $r(hip)$ according to a random mapping table. And $hip$ will be restored from $r(hip)$ according to this table too. After restoring $hip$, $sip$ will be acquired by $f^{-1}(hip)$. $RS$ has a high random because it uses IP mangling and random mapping table. But the mapping table let $RS$ generate more than one candidate hosts from $r(hip)$. 

DCDS, proposed by Wang et al. \cite{HSD:ADataStreamingMethodMonitorHostConnectionDegreeHighSpeed}, hashes $sip$ to $r(sip)$ by CRT. It restores $sip$ by solving a sequence of concurrence equations. Because the number of columns in each row is a prime different from each other, DCDS could restore $sip$ accurately. But the solving the sequence of concurrence equations requires great operations which influences the restoring speed.

In order to speed up the reversing procedure speed, VBF, proposed by Liu \cite{HSD:DetectionSuperpointsVectorBloomFilter}, hashes $sip$ to $r(sip)$ from the IP address directly. Every $r_i(sip) \in r(sip)$ is 12 bits of $sip$. By concatenating sub bits of $r(sip)$, $sip$ could be recovered successfully. But the randomness of $r(sip)$ is weak because the distribution of appearing IP addresses are not uniform. Low randomness decreases the accuracy of VBF.

Motivated by the idea of IP mangling \cite{RSL2012:TheLogLogCountingReversibleSketchADistributedArchitectureForDetectingAnomaliesBackboneNetworks} and bits concatenation \cite{HSD:DetectionSuperpointsVectorBloomFilter}, this paper proposed a high randomness and fast speed reversible hash functions group, denoted as randomness reversible hash function scheme RRH. RRH has higher randomness than VBF and faster speed than $RS$. Based on RRH, we detect super points and estimate their cardinalities more efficiently.

\section{Cardinality estimation by asynchronous time stamp vector}
LE is a vector of g bits. If a bit is visited by some hosts in a discrete time window, it will be set to 1. But a bit only has two values which limit its application in the sliding time window. The key step to estimate cardinality under sliding time window is to determine if a counter is active in the current time window. In this section, we will introduce a novel estimator, asynchronous time stamp vector $ATV$, to solve this problem.

$ATV$ is derived from LE by replacing every bit with a novel counter, asynchronous timestamp $AT$. Suppose a sliding time window contains k time slices at most. $AT$ is a counter containing $log_2(2*k+1)$ bits. It can represent $2*k+1$ different values. Let value "2*k" represent the inactive state of an $AT$. When an $AT$ equals to $2*k$, it is inactive. When the value of a $AT$ is smaller than $2*k$, the active state of this $AT$ should be determined accord to asynchronous current time stamp($ACT$). Every $AT$ is associated with an $ACT$. $ACT$ is an integer ranging from 0 to 2*k-1. When the time window sliding, $ACT$ will increase itself by 1. When $ACT$ reaches to $2*k$, it loops to 0. $AT$ has four operations: $InitAT$, $SetAT$, $checkAT$ and $preserveAT$. Suppose "$at$" is a $AT$ and "$act$" is its $ACT$. Let $Value(at)$ represent the value of "$at$".
\begin{enumerate}
\item $InitAT(at)$: set the value of "$at$" to 2*k. This operation initializes an $AT$ at the beginning. \item $SetAT(at)$: set the value of "$at$" to "$act$". When an $AT$ is mapped by some hosts, its value will be set to its $ACT$.
\item $checkAT(at, k')$: return if "$at$" is active in the latest $k'$ time slices. $k'$ is a positive integer no bigger than k. This operation is used to determine if an $AT$ is active and its detailed process is shown in algorithm \ref{alg-checkAT_at_k}.
\item $preserveAT(at)$: Update "$at$" at the beginning of every time slice. This operation signs inactive $AT$ in the new time slice. Algorithm \ref{alg-preserveAT_at} shows how this operation works.
\end{enumerate}

\begin{algorithm}         
\caption{checkAT}  
\label{alg-checkAT_at_k}  
\begin{algorithmic}[1]
\Require {
 Asynchronous timestamp $at$,
 Time slices number $k'$} 
\Ensure{ActiveState} 
\State $act \Leftarrow $ the $ACT$ of $at$
\If{$at == 2*k$}
\State Return False
\EndIf
\State $dis \Leftarrow (act+2*k-Value(at)) mod 2*k $ \label{line-disCal-alg-checkAT_at_k}
\If{$dis \leq k'-1$}
\State Return Ture
\Else
\State Return False
\EndIf
\end{algorithmic}
\end{algorithm}

\begin{algorithm}                       
\caption{preserveAT}          
\label{alg-preserveAT_at}                            
\begin{algorithmic}[1]                  
\Require {
 Asynchronous timestamp $at$,
 Time slices number $k'$} 
\State $act \Leftarrow $ the $ACT$ of $at$

\If{$(act\ mod\ k) !=0$}
\State Return
\EndIf
\If{$act == 0$}
\If{$0 \leq Value(at)\leq k$}
\State $Value(at) \Leftarrow 2*k$
\EndIf
\EndIf
\If{$act == k$}
\If{$k \leq Value(at) \leq 2*k-1$}
\State $Value(at) \Leftarrow 2*k$
\EndIf
\If{$ Value(at) == 0$}
\State $Value(at) \Leftarrow 2*k$
\EndIf
\EndIf
\end{algorithmic}
\end{algorithm}
Algorithm \ref{alg-checkAT_at_k} checks if "$at$" is active by calculating its distance with its $ACT$ at line \ref{line-disCal-alg-checkAT_at_k}. In order to saving memory, $ACT$ only has $2*k$ different values. When reaches to $2*k$, $ACT$ will roll back to 0. $AT$ stores its $ACT$ of the latest time slice that it is visited. For a time window containing $k'$ time slices where $1 \leq k' \leq k$, $AT$ has two states: inactive and active. "Active" means this $AT$ is visited by some hosts in the nearest $k'$ time slices. So $AT$ could work well under any sliding time window containing no more than $k$ time slices.

\begin{figure}[!ht]
\centering
\includegraphics[width=0.47\textwidth]{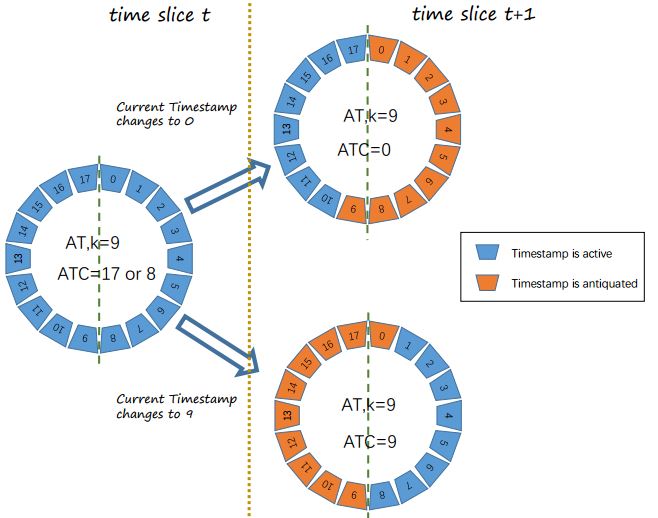}
\caption{Asynchronous timestamp}
\label{fig_asynTS_updated_everyK}
\end{figure}

$AT$ has only $2*k+1$ different values. But the sliding time window keeps moving forward permanently. So the state of $AT$ should be checked periodically at the beginning of each time slice to be labeled inactive when the distance is bigger than k. $AT$ contains one more bit than aging counter ($AC$) \cite{Iwqos2017:ACE:PerflowCountingForBigNetworkDataStreamOverSlidingWindows} which contains $log_2(k+1)$ bits. This additional bit let $AT$ have more k values than that of $AC$.
The distance calculated in line \ref{line-disCal-alg-checkAT_at_k} of algorithm \ref{alg-checkAT_at_k} could be as bigger as 2*k. When the distance of a $AT$ is bigger than $k$, $AT$ is inactive. In another word, the $AT$ state can be checked every k time slices while the state of $AC$ must be checked every time slice. Figure \ref{fig_asynTS_updated_everyK} gives an example of $AT$ with k=9.

The numbers around the circle are the values that this $AT$ could be. At a time slice, $AT$'s value is one of them. When the time window slides to time slice t+1, some numbers which have a long distance with the new $ACT$ will be changed to 18, the number that represents the inactive state. If the new $ACT$ in time slice t+1 is 0, numbers between 0 and 9 are set to 18 because their distance with 0 is bigger than $k$ at the end of time slice t+1. And when $ACT$ is 9, number 0 and numbers between 9 and 17 are set to 18. There is only one checking process every 9 time slices for this $AT$. 

If there are 2*k $AT$ and their $ACT$ are different from each other, the average checking process is O(1/k) for every $AT$ at the beginning of a time slice. Motivated by this idea, we propose the asynchronous time stamp vector $ATV$ to estimate host cardinality under sliding time window.

\begin{figure}[!ht]
\centering
\includegraphics[width=0.47\textwidth]{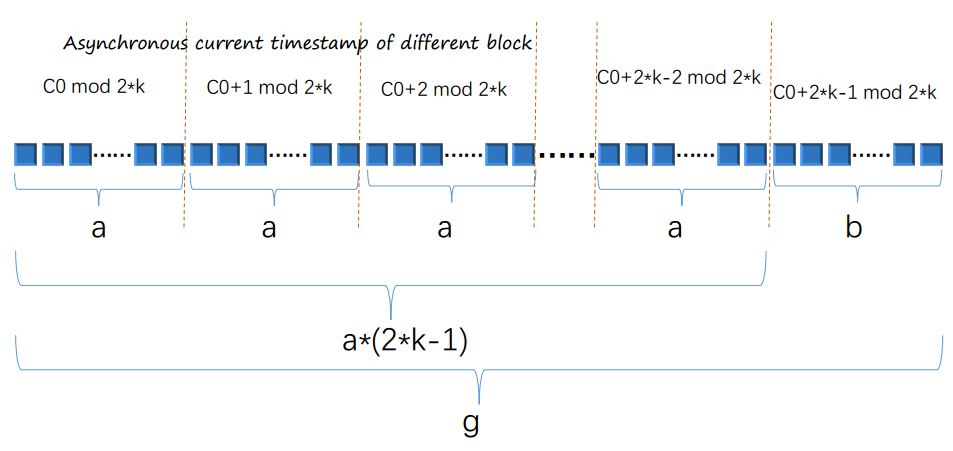}
\caption{Asynchronous timestamp vector}
\label{fig_asynTSV_group_set}
\end{figure}

$ATV$ consists of g $AT$ as shown in figure \ref{fig_asynTSV_group_set}. The g $AT$ are divided into 2*k blocks: the number of $AT$ in every of the first 2*k-1 blocks is $a$ and the number of $AT$ in the last block is $b$. $a$ and $b$ are integers and $a*(2*k-1)+b=g$. All $AT$ in a block has the same $ACT$. So a block is assigned an $ACT$. The $ACT$ of different blocks are different from each other. We only need to maintain the $ACT$ of the first block, written as $C0$. The $ACT$ of the rest 2*k-1 blocks could be acquired by $(C0+i)mod(2*k)$ where $1 \leq I \leq 2*k-1$. At the beginning of a time slice, only $AT$ in two blocks need to apply $preserveAT$ operation. Suppose $b=a=g/(2*k)$. Then the number of $AT$ of two blocks are $g/k$, and the preserving complexity of $ATV$ is only $O(g/k)$.

$ATV$ replaces every bit in $LE$ with an $AT$. Inactive $AT$ is equivalent to `0' bit and active $AT$ to `1' bit. Let $IPpair(aip, t, k)$ $=\{<aip, bip_1>,$ $<aip, bip_2>,$ $\cdots, <aip, bip_n>\}$ represent the IP pair stream extracted from packets in time slice $t_1$ and its next k-1 time slices. For every $<aip, bip> \in IPpair(aip, t, k)$, $bip$ is mapped to a AT randomly by hash function $BH(bip)$. $BH$ maps $bip$ to a value between [0, g-1]. If a AT is visited by an IP pair in a time slice, this AT will be set by $SetAT$ operation. Let $BIdx(aip, t, k)=$ $\{BH(bip)|$ $<aip, bip> \in IPpair(aip, t, k)\}$. For every $i \in BIdx(aip, t, k)$, $AT[i]$ is active at the end of last time slices of $W(t, k)$. According to equation \ref{eq_linearEstimator}, $|BIdx(aip, t, k)|$ is expected to be $g-g*e^{-(|OP(aip,t,k)|)/g}$ where $|BIdx(aip, t, k)|$ is the number of distinct elements in $BIdx(aip, t, k)$. 

The k' weight of $ATV$ is the number of active $AT$ in $ATV$, written as $|ATV|^{k'}$. $|ATV|^{k'}$ is used to estimate $|BIdx(aip, t, k)|$. Equation \ref{eq_ATV_est_cardinality} calculates the cardinality of aip by $|ATV|^{k'}$. $ATV$ is similar to LE and their diversities are listed in table \ref{tbl_ATV_LE_diversities}.

\begin{equation}
\label{eq_ATV_est_cardinality}
|OP(aip,t,k')|=-g*ln(\frac{g-|ATV|^{k'}}{g})
\end{equation}

\begin{table}
\centering
\caption{ATV vs. LE}
\label{tbl_ATV_LE_diversities}
\begin{tabular}{c}                                                                                                                                                                                                                           
\centering
\includegraphics[width=0.45\textwidth]{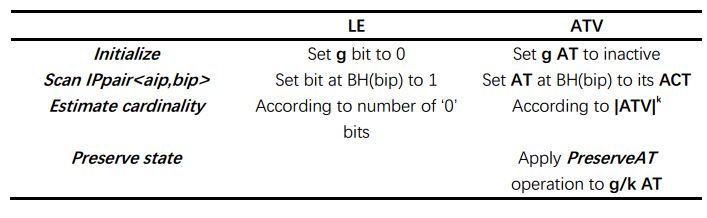}
\end{tabular}
\end{table}

$ATV$ has the same accuracy as LE. By preserving the state of every $AT$, $ATV$ can estimate cardinality in the nearest k' time slices where $k'\leq k$. In this paper, we use $ATV$ to detect super points and estimate their cardinalities under sliding time window.

\section{Super point cardinality estimation under sliding time window}
ATV estimates a host's cardinality efficiently under sliding time window. But there are millions of host in the network, and it's not reasonable to allocate an ATV for each of them because of the following two reasons:
\begin{enumerate}
\item High memory requirement. A core network always contains millions of hosts. But most of these hosts have low cardinality. Allocating an ATV containing thousands of $AT$ will waste lots of memory.

\item Frequent memory access. IP addresses of hosts are widely distributed between 0 and $2^{32}-1$, especially for IP addresses of BNet. How to store and access these randomly hosts efficiently is a hard task. No matter where these IP addresses are stored, in a list or hash table, we have to spend much time in memory accession.
\end{enumerate}
To overcome these problems, we design an ATV sharing structure, Asynchronous Timestamp Vector Cube (ATVC), which can use a fixed number of ATV to detect super points and estimate different hosts' cardinalities. ATVC is a three-dimension structure which contains $2^c*r*2^u$ $ATV$. The x dimension contains $2^c$ columns and y dimension contains $r$ rows. The set of ATV having the same z dimension is called as a frame. The z dimension contains $2^u$ frames.

ATVC has the following attributes:
\begin{enumerate}
\item For an $aip\in ANet$, there are r ATV in ATVC relating with it. By a random reversible hash function(RRH), it hashes a IP address to $r$ three-tuples RRH(aip)={$<x_0,0,z>,<x_1,1,z>,…,<x_{r-1},r-1,z>$} ;
\item Given RRH(aip), we could restore aip;
\item If $aip$ is a super point, we can acquire $RRH(aip)$ from ATVC directly;
\item For a host aip, at the end of every time slice, its cardinality could be estimated from ATVC;
\end{enumerate}
These previous attributes make sure that ATVC could estimate the cardinality of super point successfully. In this section will introduce how ATVC works in detail.

\subsection{Packets scanning}
\begin{figure}[!ht]
\centering
\includegraphics[width=0.47\textwidth]{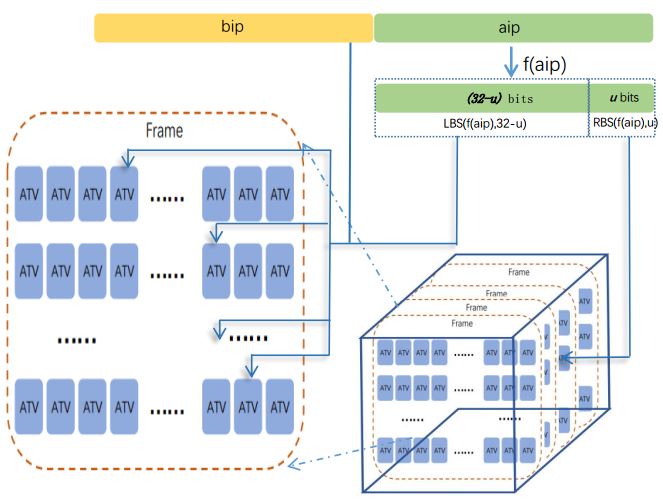}
\caption{ATV cube structure}
\label{fig_TSVC_positionMap}
\end{figure}

ATVC scans all packets in a time slice and detects super points at the end of the time slice. IPpair <aip,bip> extracted from every packet is all that ATVC needs. ATVC maps an IP pair to $r$ ATV and updates these ATV with $bip$. How to locate these ATV is the key step. RRH solves this task. RRH first mangling $aip$ to another random value $f(aip)=A*aip\ mod\ p$ where $A$ is a random positive integer no more than $2^{32}$. $``f"$ is a one-to-one mapping and $aip$ could be regain by $f^{-1}(f(aip))=A^{-1}*f(aip)\ mod\ p$ where $A^{-1}*A=1\ mod\ p$. The high randomness of RRH comes from the mangling IP process. ATV positions are acquired by extracting sub bits of $f(aip)$ as shown in figure \ref{fig_TSVC_positionMap}.

The frame index is determined by the rightest u bits of $f(aip)$, denoted by $Z(aip)$. Let $RBS(x,n)$ represent the rightest n bits of x and $LBS(x,n)$ represent the leftist n bits of x. Then $Z(aip)=RBS(f(aip),u)$. Let $LBS(x,n)[i]$ represent the $i$th bit, $LBS(x,n)[i:j]$ represent the sub bit set from $LBS(x,n)[i]$ to $LBS(x,n)[j]$. The columns indexes of these r rows are acquired by extracting $c$ bits from $LBS(f(aip), 32-u)$. Let $CIdx(aip)=\{x_0,$ $x_1, x_2,$ $\cdots, x_{r-1}\}$ represent these columns indexes and $x_i[j]$ represent the $j$th bit of $x_i$. We denote these $r$ column indexes as column tuple. In order to restore $LBS(f(aip), 32-u)$ , these hashing process has the following two attributes:
\begin{enumerate}
\item Completeness. For every $i \in [0,31-u]$, there exist at least a $j \in [0, r-1]$ and a $n \in [0, c-1]$ that $LBS(f(aip), 32-u)[i]=x_j[n]$. It means that $LBS(f(aip),$ $32-u)$ could be acquired by collecting its bits from different column indexes.
\item Redundancy. For some $i \in [0,31-u]$, there exist at least two $j1, j2 \in [0, r-1]$ and two $n1, n2 \in [0, c-1]$ that $LBS(f(aip), 32-u)[i]=x_{j1}[n1]=x_{j2}[n2]$ and $j1 \neq j2$. 
\end{enumerate}
It means that some bits of $LBS(f(aip), 32-u)$ appear in two or more column indexes.
With the second attribute, not every column tuple could restore a valid left bit set. This $i$ in the second attribute is called duplicate position. By checking if these bits in different column indexes corresponding to duplicate position are the same, we remove these column tuples that not come from a left bit set of some host.

We can use a mapping table to hash every bit of $LBS(f(aip),$ $32-u)$ to bits in different column indexes. But the table looking up process is slow. We adopt a new method that acquires these column indexes by only bit-shifting and bit-extracting operations.

\begin{figure}[!ht]
\centering
\includegraphics[width=0.47\textwidth]{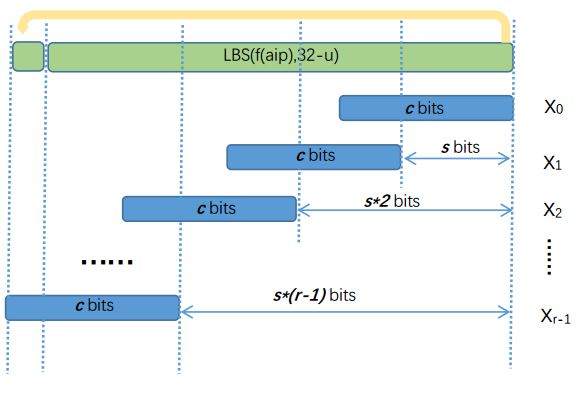}
\caption{Column index calculation}
\label{fig_ATVC_Colidx}
\end{figure}

We set $x_i$ to be successive $c$ bits of $LSB(f(aip), 32-u)$ starting from $s*i$ where $s$ is a positive integer, $1 \leq s \leq c$ and $0 \leq i \leq r-1$. For the completeness attribute, $c+s*(r-1) \geq 31-u$. If the bit position in $LSB(f(aip), 32-u)$ is bigger than 31-u, it will loop to starting from 0 as shown in figure \ref{fig_ATVC_Colidx}.

For example, we set $c=14$, $r=4$, $s=6$ and $u=3$. $x_0[0:11]=LBS(f(aip), 29)[0:11]$, $x_1[0:11]=LBS(f(aip), 29)[6:17]$. $LBS(f(aip), 29)[6:11]$ appears in $x_0$ and $x_1$ at the same time. So bit indexes from 5 to 11 are duplicate positions. For the column index $x_3$ of the last row, $x_3[0:10]=LBS(f(aip), 29)[18:28]$. The last bit of $x_3$ maps to the $29$th bit of $LBS(f(a), 29)$. This position is bigger than 28, and it will loop to 0. So $x_3[11]=LBS(f(aip),29)[0]$. Bit index 0 in $LBS(f(aip), 29)$ is also a duplicate position because it is mapped to $x_0$ and $x_3$ at the same time.

Every bit of $LBS(f(aip), 32-u)$ appears in one or several column indexes. By concatenating sub bits of these column indexes, $LBS(f(aip), 32-u)$ could be recovered successfully. Because $f(aip)$ is high random, these column indexes in $CIdx(aip)$ are high random too. 

After acquiring $CIdx(aip)$, we get $RRH(aip)=\{<x_0, 0,$ $Z(aip)>$, $<x_1, 1, Z(aip)>$, $<x_2, 2, Z(aip)>,\cdots$, $<x_{r-1}, r-1, Z(aip)>\}$. IP pair <aip, bip> will be mapped to ATV determined by $RRH(aip)$. An $AT$ in every of these $ATV$ selected by $BH(bip)$ is set to its asynchronous current time stamp where $BH$ is a hash function mapping bip to a random value between 0 and g-1. Algorithm \ref{alg-scanIPpair} shows how to scan every IP pair.
\begin{algorithm}                       
\caption{Scan IP pair}          
\label{alg-scanIPpair}                            
\begin{algorithmic}[1]                  
    \Require {IP pair <aip, bip>
              }
\For{$<x,y,z> \in RRH(aip)$}
\State $atv \Leftarrow ATVC[x,y,z]$
\State $i \Leftarrow BH(bip)$
\State $SetAT(atv[i])$
\EndFor
\end{algorithmic}
\end{algorithm}
All IP pairs extracting from packets in a time slice will be processed by this way. After scanning all packets in a time slice, super point and their cardinalities could be acquired from ATVC.

\subsection{Super points restoring}
Suppose aip is a super point in W(t,k'). If we want to restore it from ATVC at the end of the last time slice, we should know RHH(aip). According to the definition of super point, these ATV selected by RHH(aip) will give an estimating cardinality no smaller than $\theta$. By equation \ref{eq_ATV_est_cardinality}, the number of active $AT$ in every of these $ATV$ is no smaller than $|ATV|^{k'}_{\theta}=g-g*e^{-\frac{\theta}{g}}$. We call ATV whose $|ATV|^{k'}$ is no smaller $|ATV|^{k'}_{\theta}$ as super ATV. Let $SA(i, j)$ represent the set of super ATV in the $i$th row of $j$th frame. We first find all $SA(i,j)$ where $0 \leq i \leq r-1$ and $0 \leq j \leq 2^u-1$, then test super ATV frame by frame as shown in algorithm \ref{alg_restoreSIP_from_colidx}.

\begin{algorithm}                       
\caption{Restore candidate super point}          
\label{alg_restoreSIP_from_colidx}                            
\begin{algorithmic}[1]            
    \Require {SA(i, j), $i \in [0, r-1]$, $j \in [0,u-1]$
              }
    \Ensure {candidate super point mangling list CSIP }     
\For{$ j in [0, u-1]$}
\For {$\{<x_0, x_1,$ $\cdots, x_{r-1}>\}$ $\in$ $\{<SA(0,j), SA(1,j),$ $\cdots, SA(r-1,j)>\}$} \label{alg_line_colTupleSelect}
 \For {$i \in [0, r-1]$} \label{alg_line_colTupleCheck_start}
  \If {the duplicate position bits in $x_i$ not equal to that in $x_{(i+1)mod\ r}$}
  \State Continue
  \EndIf
\EndFor \label{alg_line_colTupleCheck_end}
\State $LBS \Leftarrow$ concatenate sub bits of $<x_0, x_1, \cdots, x_{r-1}>$
\State $RBS \Leftarrow j$
\State $randip \Leftarrow concatenate$ LBS with RBS
\State $sip \Leftarrow f^{-1}(randip)$
\State Insert sip into CSIP
\EndFor
\State Return CSIP
\EndFor
\end{algorithmic}
\end{algorithm}
For a frame in ATVC, line \ref{alg_line_colTupleSelect} of algorithm \ref{alg_restoreSIP_from_colidx} generates a candidate column tuple by selecting a super ATV from each row. From line \ref{alg_line_colTupleCheck_start} to \ref{alg_line_colTupleCheck_end}, the candidate column tuple is checked by bits in duplicate position. If a column tuple passes this checking, left (32-u) bits of a candidate host $aip$'s $f(aip)$ could be restored from it. Concatenated with frame index $j$ in the right, $f(aip)$ will be restored. By $f^{-1}$ function, super points $aip$ could be restored and inserted into $CSIP$. Cardinality of these candidate super points in $CSIP$ could be acquired from their corresponding $ATV$ as described in the following part.

\subsection{Cardinality Estimation}
ATVC uses fix number of ATV, $2^c*r*2^u$, to estimate the cardinalities of all hosts in ANet. It causes that a ATV will record more than one hosts' cardinalities and the result will be over estimating. To reduce the influence, r different ATV will be used together to record a host's cardinality. A super point cardinality will be estimated from these ATV. Let ATV[x,y,z] represent the ATV in the $x$th row, $j$th column of the $z$th frame. For a host $aip$, its ATV index is determined by $RHH(aip)$. Let $ATV(aip)=\{ATV[x,y,z]|$ $<x,y,z> \in RHH(aip)\}$ represent the set of ATV corresponding to $aip$. According to equation \ref{tbl_ATV_LE_diversities}, if we want to estimate the cardinality, we should calculate $|ATV|^{k'}$ as the estimation of $|BIdx(aip, t, k)|$. For every $i \in BIdx(aip, t, k)$, the $i$th AT of all ATV in $ATV(aip)$ are active. $|ATV|^{k'}$ could be estimated by calculating the number of AT that are active in all $ATV(aip)$, denoted by $NAT(aip)$. Algorithm \ref{alg-getWeightOfATV_for_cardinalityEst} shows how to acquire $ATV(aip)$ from ATVC.
\begin{algorithm}                       
\caption{Calculate the number of active AT}          
\label{alg-getWeightOfATV_for_cardinalityEst}                          
\begin{algorithmic}[1]                 
\Require {candidate super point $aip$,Time slices number k'
              }
\Ensure {active AT number $NAT(aip)$ }  

\State $NAT(aip)\Leftarrow 0$
\State $\{<x_0, 0, Z(aip)>,$ $<x_1, 1, Z(aip)>,\cdots,$ $<x_{r-1}, r-1, Z(aip)>\}\Leftarrow RRH(aip)$   
\For{$ i \in [0, g-1]$}
\State ifActive $\Leftarrow$ True
\For {$j \in [0, r-1]$}  
  \If { the $i$th AT of $ATV[x_j, j, Z(aip)]$ is not active }
     \State ifActive $\Leftarrow$ False
\State Break
  \EndIf
\EndFor  
\If {ifActive equal True}
\State $NAT(aip)\Leftarrow NAT(aip)+1$
\EndIf
\EndFor
\State Return $NAT(aip)$ 
\end{algorithmic}
\end{algorithm}

But when there are many distinct IP pairs in a time window, $NAT(aip)$ may be bigger than $|BIdx(aip, t, k)|$ caused by other hosts. Estimating the bias and removing them from $ATV(aip)$ will improve the accuracy of cardinality estimation.

Let $|AAT(k', i, j)|$ represent the number of active AT, judged by $checkAT(at, k')$ in the $i$th row of $j$th frame. The probability that a AT in the $i$th row of $j$th frame is set by some IP pair in $W(t, k')$ is $P_{i,j}^{k'}= \frac{|AAT(k', i, j)|}{g*2^c}$. $|AAT(k', i, j)|$ is acquired by scanning every ATV in the $i$th row of $j$th frame. The probability that $r$ AT in different rows of $j$th frame are all active is $UP_{k'}^{j}=\prod_{i=0}^{r-1}{P_{i,j}^{k'}}$. $|BIdx(aip, t, k)|$ is the expected number of active AT that set by aip. The rest $g-|BIdx(aip, t, k)|$ AT have probability $UP_{k'}^{Z(aip)}$ to be set to active by other hosts. The number of false active AT is expected to be $UP_{k'}^{Z(aip)}*(g-|BIdx(aip, t, k)|)$. $|NAT(aip)|$ is the sum of $|BIdx(aip, t, k)|$ and the number of false active AT as shown in equation \ref{eq_aip_nat_bidx_falseN}.
\begin{equation}
\label{eq_aip_nat_bidx_falseN}
|NAT(aip)|= |BIdx(aip, t, k)| + UP_{k'}^{Z(aip)}*(g-|BIdx(aip, t, k)|)
\end{equation}
Because $|BIdx(aip, t, k)|$ is expected to be $g$ $-$ $g*e^{-(|OP(aip,t,k)|)/g}$, we have the following equation to estimate aip's cardinality.
\begin{equation}
\label{eq_cardinality_est_after_modify}
|OP(aip, t, k')|=-g*ln\frac{g-|NAT(aip)|}{g*(1- UP_{k'}^{Z(aip)}})
\end{equation}
 
Equation \ref{eq_cardinality_est_after_modify} gives a more accurate estimation by removing the bias from $NAT(aip)$. The cardinality of every host in the candidate super point list will be estimated in this way. 

\subsection{Deploy on GPU}
In a high-speed network, such as 40 Gb/s, there are millions of packets passing through the edge of the network. To scan so many packets in real time requires plenty of computing resource. CPU is one of the most general computing part, and each core of it is very powerful to deal with complex tasks running different instructions. Though a core in CPU is powerful, its price is very high. If we want to use hundreds of CPU cores to deal with high-speed traffic parallel, we have to generate a cluster with several CPUs. The cost of the cluster will be increased with its scale. Graphics processing unit (GPU) is one of the most popular parallel computing platforms in recent years. GPU contains hundreds of processing unit in a chip, much more than that CPU has. For these tasks that have no data accessing conflict and processing different data with the same instructions (SIMD), GPU can acquire a high-speed up\cite{PD2013:BenchmarkingOfCommunicationTechniquesForGPUs}\cite{PD2013:GeneratingDataTransfersForDistributedGPUParallelPrograms}. 

For ASSE, every packet is processed by algorithm \ref{alg-scanIPpair}. Algorithm \ref{alg-scanIPpair} just sets some AT to its asynchronous current timestamp and every AT could be set by different threads concurrently without introducing any mistakes. GPU only accesses its graphic memory directly. So we put ATVC on GPU memory. ATVC scans every IP pair, and every IP pair must be copied to GPU memory. Copying IP pair one by one is inefficient because a data transmission routine between server memory and GPU memory requires additional starting and ending operations. To improve the efficiency, we allocate two buffers which can store thousands of IP pairs on both GPU memory and server memory. When the buffer on the server side is full, thousands of IP pairs in it will be copied to the buffer on GPU side. After receiving these IP pairs, thousands of GPU threads running algorithm \ref{alg-scanIPpair} are launching together to deal with these IP pairs. 

Not only IP pairs scanning but also super point restoring is running on GPU. At the end of a time slice, thousands of threads are launched to get super ATV in different rows. Then every candidate column tuple is assigned a thread to restore super point and estimate its cardinality. When running on GPU, ATVC estimates super points cardinalities on a core network in real time as shown in the next section. 

\section{Experiments}
To evaluate the performance of ASSE, we use real-world traffic downloaded from Caida \cite{expdata:Caida}. The experiment data are four one-hour traffics between Seattle and Chicago starting from 13:00 on different days and their average summary in a 5-minute time window are listed in table \ref{fig_expr_traffic_info}. We set Seattle as $ANet$ and Chicago as $BNet$. "$\#ANet\ IP$" and "$\#BNet\ IP$" are the average number of distinct hosts in $ANet$ and $BNet$ separately. "$\#Flow$" is the average number of distinct IP pairs. There are two parts in our experiments: super point cardinality estimation under discrete time window and super point cardinality estimation under sliding time window. In both of these parts, super point's threshold $\theta$ is set to 1024. All experiments run on a PC with GPU card Nvidia Titan XP(12 GB graphic memory).
\begin{table}
\centering
\caption{Traffic summary}
\label{fig_expr_traffic_info}
\begin{tabular}{c}                                                                                                                                                                                                                           
\centering
\includegraphics[width=0.45\textwidth]{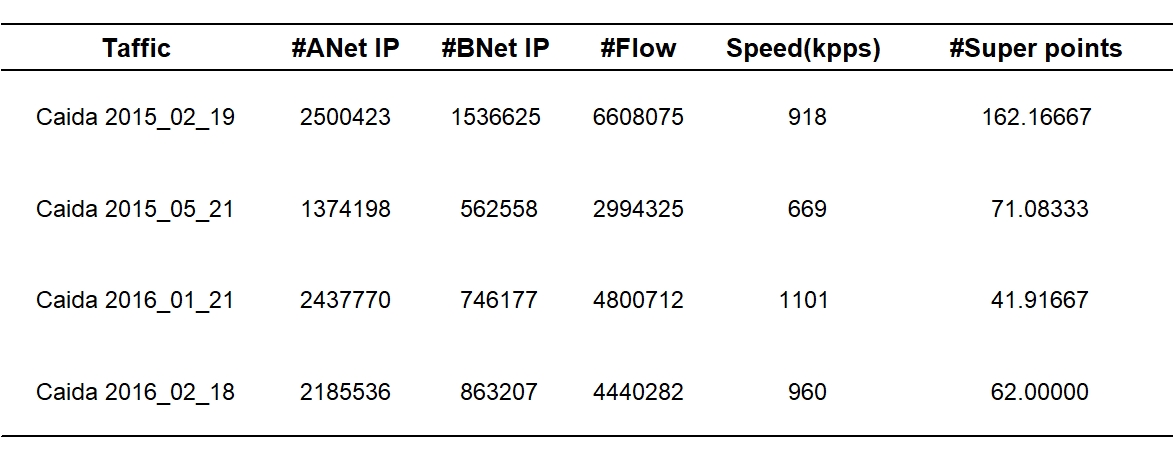}
\end{tabular}
\end{table}

\subsection{Discrete time window experiments}
We set the size of a discrete time window to 5 minutes. There are 12 discrete time windows in each traffic. In the discrete time window, AT also uses a bit to represent its state. Fig. \ref{fig_expr_compareRlt_dtw} shows the compare detection result of ASSE with DCDS\cite{HSD:ADataStreamingMethodMonitorHostConnectionDegreeHighSpeed}, VBFA\cite{hsd:CompactSpreadEstimatorSmallHighSpeedMemory} and GSE\cite{HSD:GPU:2014:AGrandSpreadEstimatorUsingGPU}. 
Accuracy is the key merit of super point detection. We measure the accuracy of super point detection algorithm according to false positive rate(FPR), false negative rate(FNR) as defined below.
\begin{definition}[FPR/FNR]
\label{def_fpr_fnr}
For traffic with N super points, an algorithm detects N' super points. In the N' detected super points, there are $N^+$ hosts which are not super points. And there are $N^-$ super points which are not detected by the algorithm. FPR means the ratio of $N^+$ to N and FNR means the ratio of $N^-$ to N.
\end{definition}
FPR may decrease with the increase of FNR. If an algorithm reports more hosts as super points, its FNR will decrease, but FPR will increase. So we use the sum of FPR and FNR, the total false rate FTR, to evaluate the accuracy of an algorithm. In the experiment, we set g=4096, r=4, c=14, and u=4. The detection results are listed in table \ref{fig_expr_compareRlt_dtw}. In table \ref{fig_expr_compareRlt_dtw}, $Cu$ is the time consumed by different algorithms for packets scanning and $Ce$ for super point restoring.
\begin{table}
\centering
\caption{Detection result comparing}
\label{fig_expr_compareRlt_dtw}
\begin{tabular}{c}                                                                                                                                                                                                                           
\centering
\includegraphics[width=0.45\textwidth]{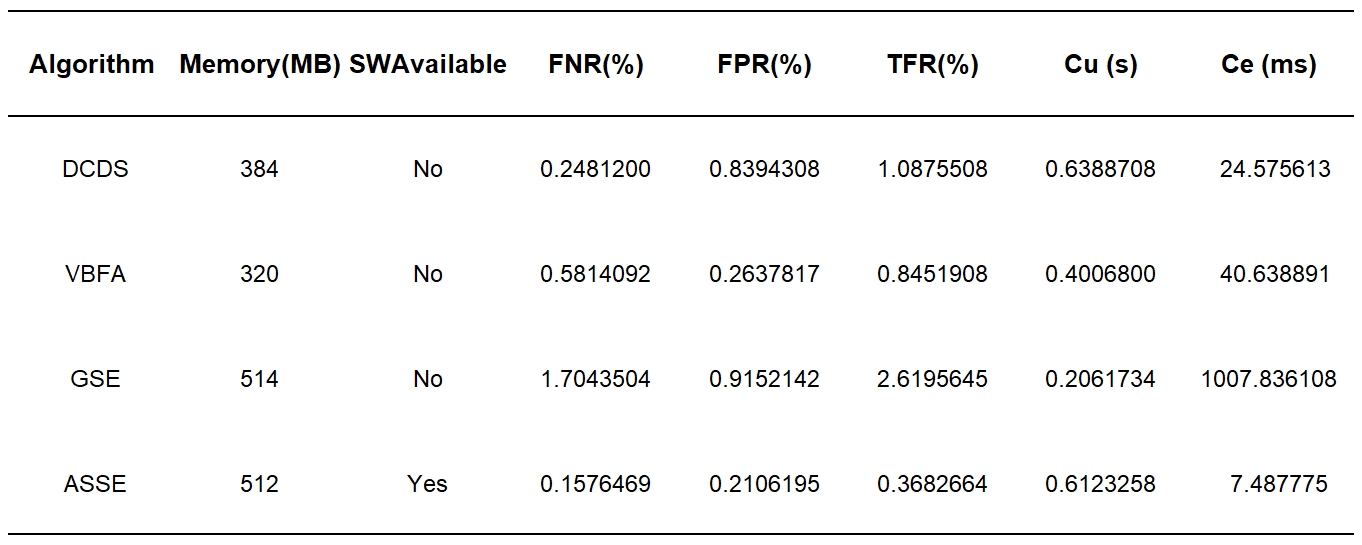}
\end{tabular}
\end{table}

GSE has a higher false rate than other algorithms because it only uses a single virtual estimator for a host and the bit of this virtual estimator is also shared by other virtual estimators. The rest algorithms use several cardinalities estimators for every host at the same time: DCDS uses three estimators, VBFA uses five estimators, and ASSE uses four estimators. For a host, its cardinality is estimated from the union of these estimators. The union of estimators reduces the influence of bits sharing and improves the estimation accuracy. Generally, the more estimators, the higher accuracy. So ASSE and VBFA have higher accuracy than DCDS. But ASSE selects estimators for a host more randomly than VBFA and it has the highest accuracy among all of these algorithms.

Because GSE only uses a logical estimator for a host, it only needs to update a bit for a packet. So GSE has the lowest packets scanning time. But GSE needs to estimate all hosts cardinality to detect super points, so it spends the most time in super point detection. DCDS maps a host to different estimators by CRT which requires another two dividing operations. But VBFA maps a host to different estimators by bits extracting which is faster than mathematical operations. So VBFA uses smaller packets scanning time than DCDS. Although ASSE maps a host to different estimators by bits extracting too, it uses the mangling IP operation to increase the randomness which also increases the processing time. As mentioned before, DCDS and VBFA map a host to three and five different estimators, they also restore a super point from these three or five estimators used by this super point. But VBFA generates much more such kind of estimators tuple for testing which let it uses much more super point restoring time than DCDS. ASSE splits traffic to different frames, and super points are also divided into different frames too. Each frame only contains a fraction of super points, and the number of candidate active column tuples reduces greatly. So ASSE has the smallest super point detection time. Super point detection time is very important under sliding time window because at the end of every time slice super point will be detected once which is more frequently than that under discrete time window.  

\begin{figure}[!ht]
\centering
\includegraphics[width=0.47\textwidth]{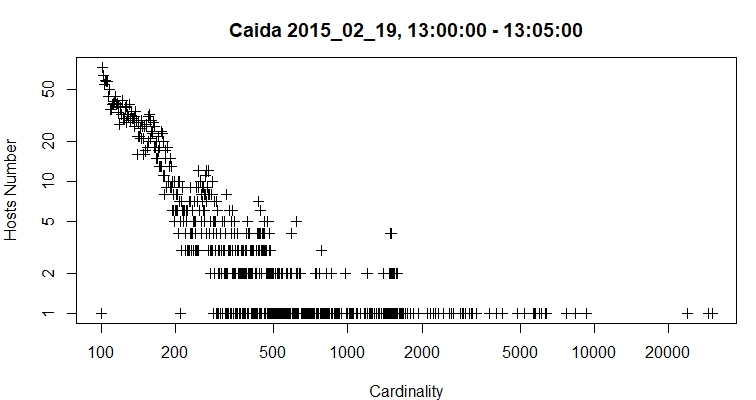}
\caption{Cardinality distribution in a time window}
\label{fig_expr_traffic_realCardinalityDistribute}
\end{figure}

\begin{figure}[!ht]
\centering
\includegraphics[width=0.47\textwidth]{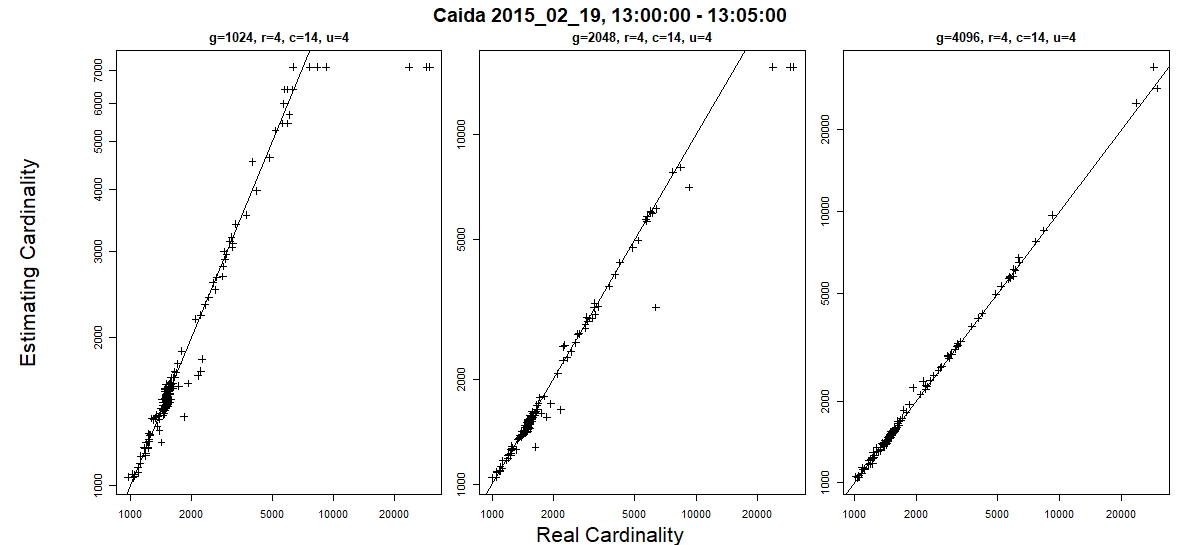}
\caption{Estimate cardinality vs. real cardinality}
\label{fig_expr_est_real_cardinality_differentPar}
\end{figure}

ASSE estimates the cardinality of every detected super point. In five minutes, most hosts' cardinalities are smaller than 5000. Fig.\ref{fig_expr_traffic_realCardinalityDistribute} shows the cardinality distribution in a 5-minute time window. The accuracy of ATV is affected by g. When g changes from 1024 to 4096, the accuracy of estimated cardinality given by ASSE increases too as shown in fig. \ref{fig_expr_est_real_cardinality_differentPar}.

When g is equal to 1024, there are 6 points whose cardinalities are bigger than 5000 not being evaluated well. But when g is set to 4096, all super points' cardinalities are estimated accurately. 

\subsection{Sliding time window experiments}
\begin{figure}[!ht]
\centering
\includegraphics[width=0.47\textwidth]{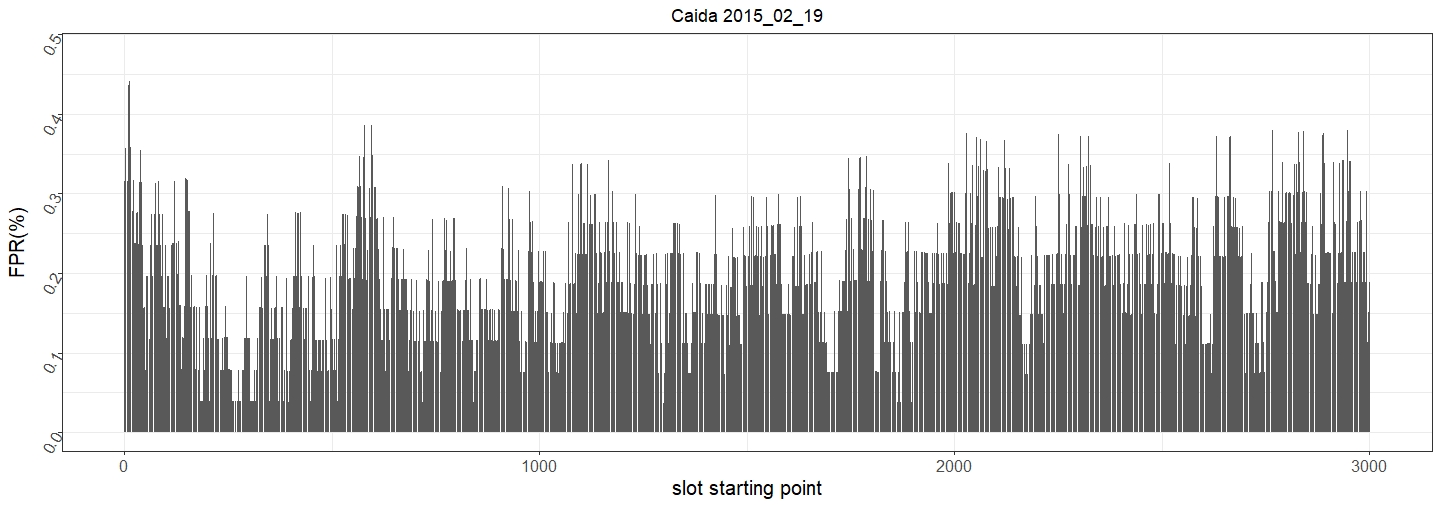}
\caption{FPR under sliding time window}
\label{fig_expr_sw_fpr}
\end{figure}

\begin{figure}[!ht]
\centering
\includegraphics[width=0.47\textwidth]{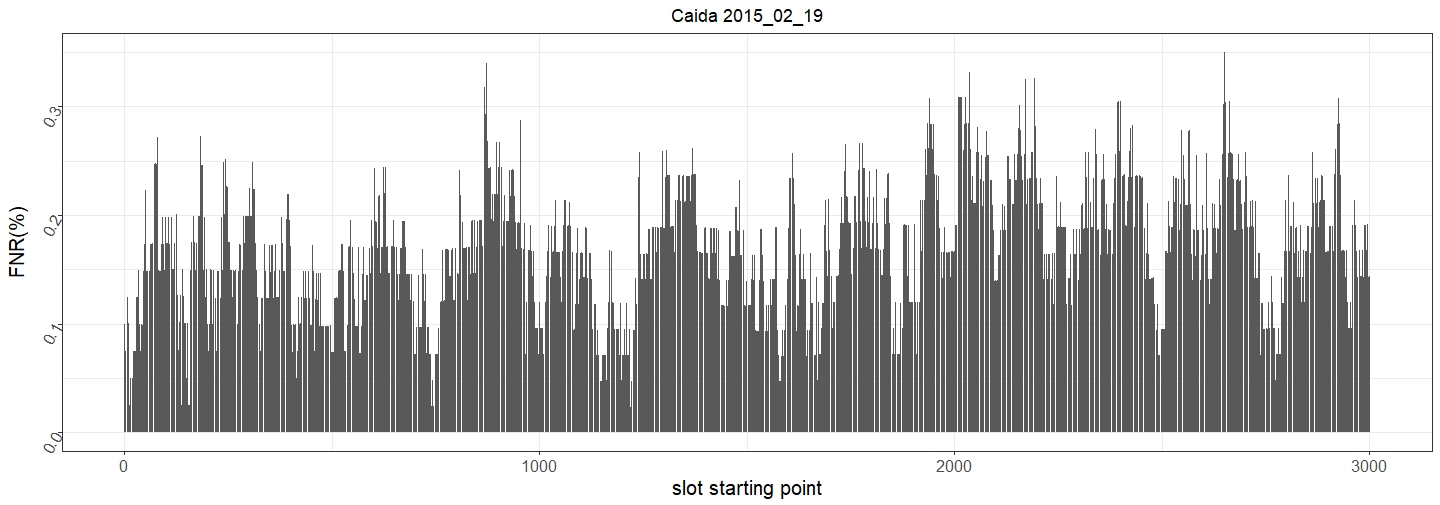}
\caption{FNR under sliding time window}
\label{fig_expr_sw_fnr}
\end{figure}

\begin{figure}[!ht]
\centering
\includegraphics[width=0.47\textwidth]{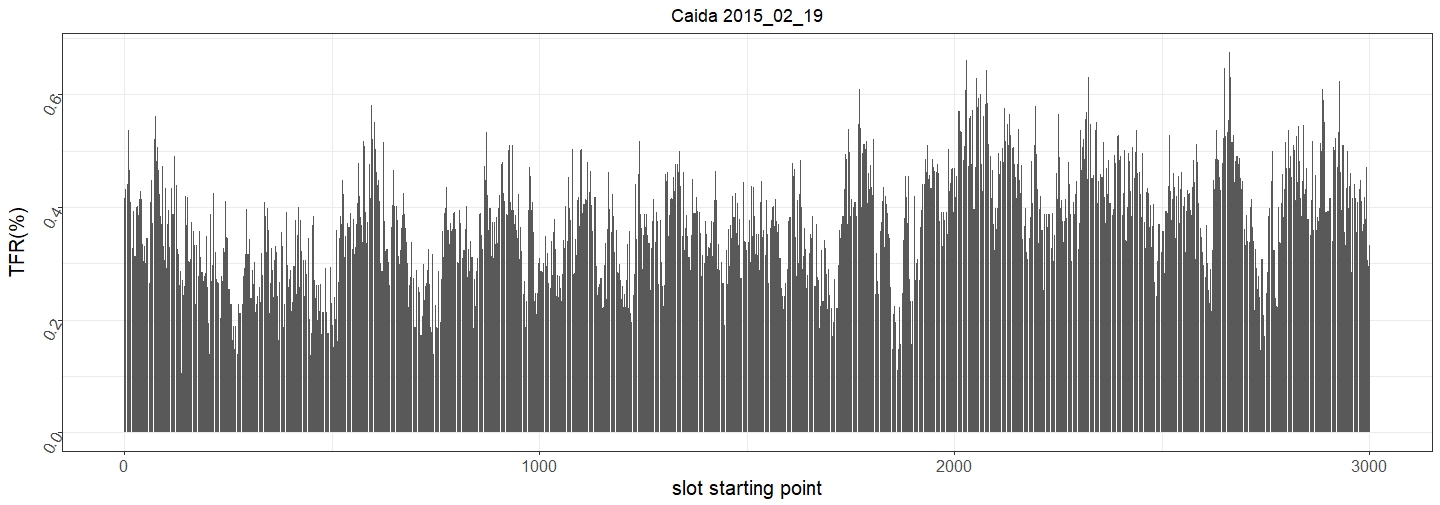}
\caption{TFR under sliding time window}
\label{fig_expr_sw_ftr}
\end{figure}

In the sliding time window experiments, a time slice is set to 1 second, k is 300. We run ASSE on the first traffic $Caida$\ $2015\_02\_09$. We let the window sliding from W(0, 300) to W(3299, 300) where the first second of this traffic is set as the first time slice. ASSE's FPR, FNR and TFR are illustrated in Fig \ref{fig_expr_sw_fpr}, \ref{fig_expr_sw_fnr} and \ref{fig_expr_sw_ftr}.

Under every sliding time window, ASSE has a low FNR, as small as $0.17\%$ on average. When FNR is small, FPR is relatively high. But the total false rate is stably small, only $0.27\%$ on average. When under sliding time window, ASSE has the similar accuracy as it has under discrete time window. This experiment proves that ASSE estimates super point cardinality successfully under sliding time window on GPU. In the sliding time window experiments, ASSE's need to preserve the states of AT at the end of every time slice. The average time consumed by this procedure is two milliseconds. The small consuming time benefits from the fact that an AT only needs to be checked every 300 time slices. In a time slice, ASSE consumes 20 million seconds for packets scanning and 15 million seconds for super point cardinality estimating on average. The total time used by ASSE in a time slice is much smaller than the size of a time slice. So ASSE can estimate the cardinalities of super points in real time under sliding time window.
\section{Conclusion}
Super point cardinality estimation under sliding time window in real time is an important topic in network research. Incremental updating and small estimating time are two special difficulties in it. ASSE proposed in this paper is a sliding time window available algorithm which can estimate super point cardinality in real time. ASSE's capability of incremental updating comes from ATV, a new estimator consisting of several asynchronous timestamps. For a sliding time with k time slices, an asynchronous time stamp only needs to be updated every k time slices. Time used for ATV preserving is reduced greatly. ASSE uses ATV cube(ATVC) to maintain all hosts cardinalities under sliding time window. By a random reversible hash function scheme, ASSE restores super points and estimate their cardinalities from ATVC. ASSE is a parallelable algorithm. When running on GPU, ASSE can estimate the cardinality of a super point from core network in real time under sliding time window.

 \iftoggle{ACM}{
\bibliographystyle{unsrt}
}

 \iftoggle{ElsJ}{
 \bibliographystyle{elsarticle-num}
}

\bibliography{..//..//ref} 

\end{document}